\documentclass[doublecol]{epl2} 
\usepackage{amsmath}
\usepackage{graphicx}
\usepackage{amsfonts}
\usepackage{amssymb}
\usepackage{amsmath}
\usepackage{color}

\newcommand{\ve}[1]{{\mathbf #1}}

\newcommand{\bra}[1]{\langle\left.{#1}\right|}
\newcommand{\ket}[1]{\left|{#1}\right.\rangle}

\title{Repulsive polarons in two-dimensional Fermi gases}

\author{Vudtiwat Ngampruetikorn\inst{1} \and Jesper Levinsen\inst{1} \and Meera M. Parish\inst{1,2}}
\shortauthor{V. Ngampruetikorn, J. Levinsen and M. M. Parish}

\institute{                    
  \inst{1} 
T.C.M. Group, Cavendish Laboratory, JJ Thomson Avenue, Cambridge CB3
  0HE, United Kingdom \\
  \inst{2} London Centre for Nanotechnology - Gordon Street, London, WC1H 0AH, United Kingdom
}

\abstract{
  We consider a single spin-down impurity atom interacting via an
  attractive, short-range potential with a spin-up Fermi sea in two
  dimensions (2D). Similarly to 3D, we show how 
  the impurity can form
  a metastable state (the ``repulsive polaron'') with energy greater
  than that of the non-interacting impurity.  Moreover, we find that
  the repulsive polaron can acquire a finite momentum for sufficiently
  weak attractive interactions.  Even though the energy of the
  repulsive polaron can become sizeable, we argue that saturated
  ferromagnetism is unfavorable in 2D because of the polaron's finite
  lifetime and small quasiparticle weight.
}

\begin{document}

\maketitle

\section{Introduction}
The Fermi gas with effectively repulsive, short-range interactions
provides a model system in which to investigate magnetic instabilities
of the Fermi liquid state.  Of particular interest is the putative
Stoner transition to itinerant ferromagnetism, which is driven purely
by the repulsive interactions.  Such a scenario can potentially be
realized and investigated with ultracold, dilute gases of
atoms~\cite{2009Sci...325.1521J}.  However, recent cold-atom
experiments in three dimensions (3D) have cast doubt on whether the
Stoner transition can ever be realized in a strongly repulsive Fermi
gas with short-range interactions~\cite{sanner2011}.  The central
issue is that strongly repulsive interactions in 3D can only be truly
short-ranged if the underlying potential is attractive. This
necessarily demands that any magnetic phase be metastable at best,
since the true ground state will involve attractively-interacting
pairs of $\uparrow$ and $\downarrow$ fermions. In 3D, the
ferromagnetic state is apparently never stable~\cite{sanner2011,Pekker2011,Zhang2011}, thus
a key question is whether this is also true in lower dimensions.

Lowering the dimensionality will strongly modify both the two-body
scattering properties and the many-body properties.  In 1D, it is in
fact permissible to have purely repulsive contact interactions, but
here the Lieb-Mattis theorem precludes any ferromagnetic ground
state~\cite{Lieb1962}. The situation is less clear in 2D, however,
where one still requires an underlying attractive potential to
generate strong, short-range repulsive interactions, but in contrast
to 3D there is always a two-body bound state and the two-body
scattering amplitude is energy dependent, even at low
energies~\cite{PhysRevB.41.327}.

Fortunately, 2D Fermi gases have recently been realized in cold-atom
experiments~\cite{Modugno2003,Gunter2005,2DFermi_expt,PhysRevLett.106.105301,sommer2011_2D,Dyke2011,Zhang:2012uq},
making it an ideal time in which to investigate ferromagnetism in
2D. Theories based on a mean-field approach currently support the
existence of itinerant ferromagnetism~\cite{Conduit:2010ve}. However,
given the importance of quantum fluctuations~\cite{Conduit:2010ve}, we
ideally want to explore limits of the problem where the theory is more
controlled. To this end, we consider the limit of extreme spin
imbalance, where we have one spin-down ``impurity'' immersed in a
Fermi sea of spin-up particles.  Here, theoretical approaches based on
simple variational wave 
functions~\cite{PhysRevA.74.063628,PhysRevLett.101.050404}
 have proved to be extremely reliable in 3D
when compared with experiments~\cite{PhysRevLett.102.230402,PhysRevLett.103.170402}
and quantum Monte Carlo simulations~\cite{PhysRevB.77.020408,PhysRevB.77.125101}. Thus
far, the ground state of the 2D impurity problem has been
theoretically investigated~\cite{Parish:2011vn,
  Zollner:2011fk,Klawunn2011},
 and it has even been suggested that recent 2D experiments have already
observed single-impurity physics~\cite{schmidt2011}.

In this Letter, we show how a $\downarrow$-impurity interacting
attractively with a $\uparrow$ Fermi sea in 2D can form a metastable
state --- the so-called ``repulsive polaron'' --- with energy greater
than that of the non-interacting impurity. Our focus is on equal
masses ($m_\uparrow = m_\downarrow$), given its relevance to current
cold-atom experiments and ferromagnetism, but we also investigate how
the repulsive polaron depends on the mass ratio
$m_\uparrow/m_\downarrow$.  As the attraction is 
decreased, we find
that the energy $E_+$ of the repulsive polaron increases and
eventually surpasses the Fermi energy $\varepsilon_F$ of the spin-up
Fermi sea.  Moreover, we find that the repulsive polaron acquires a
finite momentum for sufficiently weak attraction. 
However, its decay to the ground-state ``attractive'' polaron is also
enhanced when $E_+ \simeq \varepsilon_{F}$, which suggests that
saturated ferromagnetism is unlikely to exist in 2D, as we argue
below.

\section{Methods} 
In the following, we consider a two-component ($\uparrow$,
$\downarrow$) 2D Fermi system with short-range attractive
interactions, described by the Hamiltonian
\begin{equation}\label{Hamiltonian}
H = \sum_{\mathbf{p}\sigma} \epsilon^\sigma_\mathbf{p} c^\dagger_{\mathbf{p}\sigma} c_{\mathbf{p}\sigma}
+ \frac{g}{\Omega} \sum_\mathbf{kpq} c^\dagger_{\mathbf{k}\downarrow} c^\dagger_{\mathbf{p}\uparrow} c_{\mathbf{p+q}\uparrow} c_{\mathbf{k-q}\downarrow},
\end{equation}
where $\epsilon^\sigma_\mathbf{p} = p^2/2m_\sigma$ (with $\hbar \equiv
1$), $\Omega$ is the system area and $g$ is the bare coupling
constant describing the interspecies contact interaction. In addition
we note that, for fermions, the interaction is interspecies only since
the exclusion principle forbids intraspecies $s$-wave scattering. In
2D, the bare coupling strength can be related to the two-body binding
energy $\varepsilon_B$, which is always present for attractive
interactions in 2D, via \cite{PhysRevB.41.327}
\begin{align}
-\frac{1}{g} = \frac{1}{\Omega} \sum^\Lambda_\mathbf{p} \frac{1}{\varepsilon_B + \epsilon^\uparrow_\mathbf{p}+\epsilon^\downarrow_\mathbf{p}}. 
\end{align}
The cut-off momentum $\Lambda$ can be sent to infinity at the end of
the calculation.

To proceed, we adopt the variational polaron wavefunction introduced
in Ref.~\cite{PhysRevA.74.063628},
\begin{equation} \label{polaronwavefunction}
|P\rangle 
= \alpha_0^\mathbf{(p)} c^\dag_{\mathbf{p}\downarrow}|FS\rangle + \sum_\mathbf{kq} \alpha^\mathbf{(p)}_\mathbf{kq} c^\dag_{\mathbf{p+q-k}\downarrow} c^\dag_{\mathbf{k}\uparrow} c_{\mathbf{q}\uparrow}|FS\rangle.
\end{equation}
The first term on the \textit{r.h.s.} describes a bare impurity and
the second term takes into account one particle-hole pair excitation
of the Fermi sea, $|FS\rangle$. The polaron ground-state energy is determined by
minimising $ \langle P|H|P\rangle$ while keeping $\langle P | P\rangle
=1$.  
The variational approach can also be generalized to study metastable excited states such 
as the repulsive polaron by instead minimising the action 
$\int dt \bra{P}(i\hbar \partial_t - H)\ket{P}$ 
 ---  see, e.g.,\ Ref. \cite{PhysRevE.51.5688} for further details. 
Equivalently, one can use the 
diagrammatic method in which the interspecies interaction is treated
in the ladder approximation~\cite{Combescot:2007bh}. This yields the
expression for the self-energy:
%
\begin{equation} \label{selfenergy}
\Sigma(\mathbf{p},E) =  \sum_\mathbf{q} \left[\frac{\Omega}{g} - \sum_\mathbf{k}\frac{1}{E+i0^+ - E_\mathbf{kq;p}}\right]^{-1}.
\end{equation}
%
$E_\mathbf{kq;p} =
\epsilon^\downarrow_\mathbf{p+q-k}+\epsilon^\uparrow_\mathbf{k}-\epsilon^\uparrow_\mathbf{q}$
and $E$ measures the energy change due to the addition of an impurity
to the Fermi sea. Here, and in what follows, the hole and particle
momenta are limited to $|\mathbf{q}| < k_F$ and $|\mathbf{k}| > k_F$,
respectively, with $k_F$ the majority Fermi momentum. In principle,
one can obtain a more accurate description by allowing more
particle-hole pair excitations; however, it has been shown that one
particle-hole pair excitation yields a good estimate in 3D, as we will
discuss below.

From the self-energy (\ref{selfenergy}), one can obtain the energy of
the polaron by locating the quasiparticle pole,
\begin{equation}\label{qpole}
\varepsilon = \frac{p^2}{2m_\downarrow}+\Sigma(\mathbf{p},\varepsilon).
\end{equation}
Solutions of this equation are in general complex, the presence of an
imaginary part indicating a finite lifetime of the quasiparticle. If
the decay rate $\Gamma$ is much smaller than
$\mathfrak{Re}\{\varepsilon\}$, the quasiparticle energy is given by
\begin{equation}
E = \frac{p^2}{2m_\downarrow} + \mathfrak{Re}\{\Sigma(\mathbf{p},E)\}. \label{energy}
\end{equation}
We quantify the momentum dependence of the energy $E$ with the
effective mass $m^* \equiv 1/[\partial^2 E/\partial p^2]_{p=0}$. The
quasiparticle weight is described by the residue
\begin{equation}
Z = \left|\alpha_0^\mathbf{(p)}\right|^2 \simeq\left[ 1 -
  \frac{\partial\,
\mathfrak{Re}\{\Sigma(\mathbf{p},E')\}}{\partial E'}\right]^{-1}_{E'=E}.
\end{equation}
The approximation becomes exact in the limit $\Gamma/E\to 0$.

Equation~(\ref{energy}) yields two solutions for the energy $E$. The
negative energy solution $E_-$ corresponds to the attractive polaron
studied in Refs.~\cite{Parish:2011vn, Zollner:2011fk}.  For a weakly
interacting system, the attractive polaron is the ground state of an
impurity immersed in the Fermi sea. However, as the strength
  of the attractive interaction is
increased, a bound diatomic molecule dressed by particle-hole
excitations becomes energetically favorable \cite{Parish:2011vn}.
Additionally, a solution of Eq.~(\ref{energy}) with positive energy
$E_+$ exists as found in Ref.~\cite{schmidt2011}. The
theory of this repulsive polaron was recently elucidated in
3D~\cite{Cui:2010zr,Massignan:2011fk,PhysRevA.83.063620}.  This
metastable quasiparticle is dressed by a cloud of particle-hole pair
excitations whose effect is, in contrast to the attractive case, a
reduced majority particle density around the
impurity~\cite{Massignan:2011fk}. However, the repulsive polaron is
not the true ground state (indeed, it is not generally an eigenstate
of the system) and thus it will eventually decay.  Energy and momentum
conservation restrict the possible decay modes: The simplest are into
either an attractive polaron and a particle-hole pair or into a
molecule, two holes and one particle
\cite{PhysRevB.77.020408,PhysRevB.77.125101,Massignan:2011fk}, the
first being the main source of decay in the region of interest as we shall argue below.

The decay rate of the repulsive polaron sets the timescale at which
the physics related to this excited state may be observed.  In
general, if the quasiparticle self-energy is known exactly, the rate
may be calculated as $\Gamma =- Z_+
\mathfrak{Im}\{\Sigma(\mathbf{p},E_+)\}$. However, when the
self-energy is approximate, a reliable estimate requires the lower
lying states into which the quasiparticle can decay to be incorporated
correctly. The self-energy obtained from the wavefunction
(\ref{polaronwavefunction}) is unable to accurately capture the decay
to an attractive polaron (described by the same wavefunction) and a
particle-hole pair.  Instead, we calculate the rate of decay to the
attractive polaron following the idea of Ref.~\cite{Massignan:2011fk}:
Close to the quasiparticle poles, the repulsive ($+$) and attractive
($-$) polarons are described by Green's functions
\begin{equation}
G_\pm(\mathbf{p},\omega)\sim
\frac{Z_\pm}{\omega-E_\pm-p^2/(2m^*_\pm)}.
\label{eq:pole}
\end{equation}
The decay to the attractive polaron is dominated by processes which
correctly match the energy difference $\Delta E\equiv E_+-E_-$. Thus,
we estimate the decay rate by replacing the bare impurity propagator
in the ladder summation by the pole expansion of the attractive
polaron.  In this manner we arrive at the expression
\begin{equation}
\Gamma=-Z_+ \mathfrak{Im}\sum_\mathbf{q} \left[\frac{\Omega}{g^*}
  - \sum_\mathbf{k}\frac{Z_-}{\Delta E+i0^+ -
    E_\mathbf{kq;p}^*}\right]^{-1}.
\label{gamma}
\end{equation}
The energy $E_\mathbf{kq;p}^*$ is obtained from $E_\mathbf{kq;p}$ by
replacing $m_\downarrow$ with the effective mass of the attractive
polaron, and the coupling constant $g$ is adjusted to obtain
ultraviolet convergence\footnote{We expect the impurity to act as a
  free particle for momentum $k\gg k_F$. Instead of
  using a complicated dispersion for the attractive polaron
  (effectively letting $m^*_-$ depend on momentum), we note that the
  large $k$ behavior is cut off by a corresponding term in the
  coupling $g$ and adjust the coefficient in front of this term to
  cancel the ultraviolet divergence.}.


\begin{figure}
\centering
\includegraphics[width=1.00\linewidth]{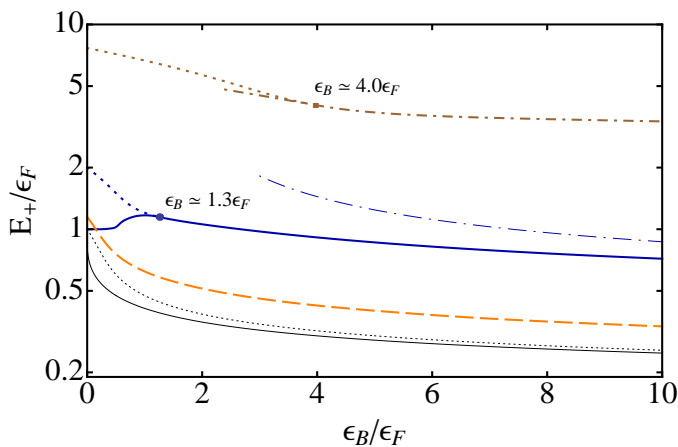}
\caption{(Color online) Energy of the repulsive polaron as a function
  of $\varepsilon_B/\varepsilon_F$ for different experimentally
  relevant mass ratios (thick): From top to bottom,
  $m_\uparrow/m_\downarrow = 40/6,\, 1,\, 6/40 $.  Where these curves
  split, the dotted lines depict the zero-momentum energy while the
  lower branch traces the minimum of the polaron dispersion.
  We also display the asymptotic result (\ref{eq:asymp}) for
    equal masses (double-dash-dotted). At the bottom, the
    thin (black) lines are the energies of an infinitely massive
  impurity as obtained from the exact solution (solid) and the ladder
  approximation (dotted).  As $\varepsilon_B/\varepsilon_F \to 0$,
  both of these approach $\varepsilon_F$. }
 \label{fig:energy}
\end{figure}

\begin{figure}
\centering
\includegraphics[width=1.00\linewidth]{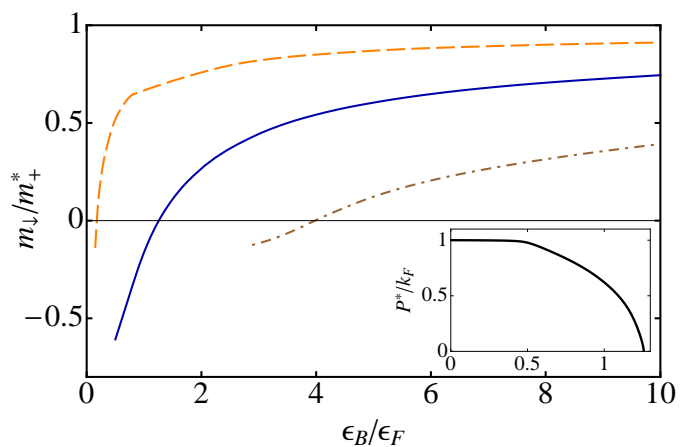}
\caption{(Color online) Inverse effective mass 
  of the repulsive polaron for mass ratios
  $m_\uparrow/m_\downarrow=40/6 \text{ (dot-dashed)}$, $1 \text{
    (solid)}$, $6/40$ (dashed).  Where $m_+^*<0$, %
  a polaron with finite momentum will be energetically favorable.
  Ref.~\cite{schmidt2011} considered the equal mass case in the regime
    $\varepsilon_B/\varepsilon_F\gtrsim3$ and their results agree with ours.
\textit{Inset}:
  ($m_\uparrow\hspace{-0.5mm}=\hspace{-0.5mm}m_\downarrow$) Momentum
  $P^*$ which minimizes the energy of the polaron.}
 \label{fig:mass}
\end{figure}

\section{Repulsive polaron}
\begin{figure}
\centering
\includegraphics[width=1.00\linewidth]{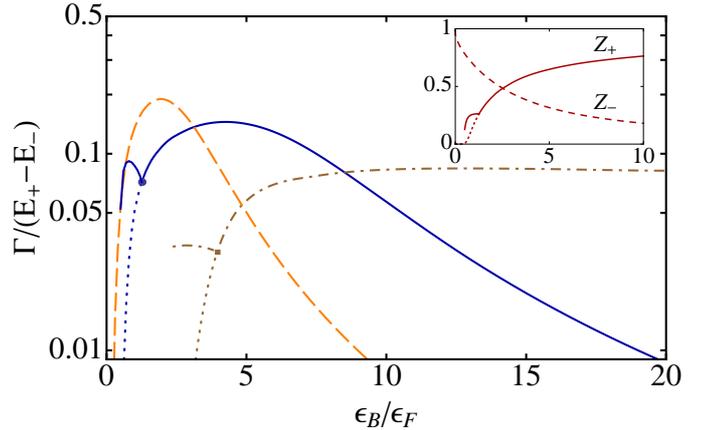}
\caption{(Color online) Decay rate $\Gamma$ of the repulsive polaron
  for mass ratios $m_\uparrow/m_\downarrow=40/6 \text{ (dot-dashed)}$,
  $1 \text{ (solid)}$, $6/40$ (dashed). Where these lines split, the
  upper and lower curves depict $\Gamma$ for the finite- and
  zero-momentum polaron, respectively. \textit{Inset}:
 ($m_\uparrow=m_\downarrow$) Quasiparticle residues. The solid line
  follows the lowest lying repulsive polaron and the dotted line
  the zero-momentum state. Ref.~\cite{schmidt2011} considered the
  zero-momentum states and our results are consistent.}
\label{fig:decay}
\end{figure}

The properties of the repulsive polaron are parameterized by the
dimensionless quantity $\varepsilon_B/\varepsilon_F$, which gives a
measure of the interaction strength. For strong attraction, 
$\varepsilon_B/\varepsilon_F \gg 1$, the repulsive polaron branch is
only weakly perturbed by the two-body bound state which sits far below
the continuum, and the repulsive polaron approaches the bare impurity
state. Consequently, when $\varepsilon_B/\varepsilon_F \to \infty$,
the energy $E_+ \to 0$, the residue $Z_+ \to 1$ and the effective mass
$m^*_+ \to m_\downarrow$ , as depicted in
Figs.~\ref{fig:energy}-\ref{fig:decay}, respectively. Indeed, in this
limit we find
\begin{equation}
 E_+/\varepsilon_F \simeq (m_\uparrow +
m_\downarrow)/[m_\downarrow \log(\varepsilon_B/\varepsilon_F)]
\label{eq:asymp}
\end{equation}
to leading order, which is consistent with previous perturbative
calculations for equal masses \cite{PhysRevB.12.125}. The two-body
decay rate $\Gamma$ is also logarithmically suppressed for
$\varepsilon_b\gg \varepsilon_F$: From Eq.~(\ref{gamma}),
$\Gamma\simeq \frac{m^*_-(m_\uparrow+m_\downarrow)^2}
{m_\downarrow^2(m_\uparrow+m^*_-)} \frac{\pi
  Z_+Z_-\varepsilon_F}{[\log(\varepsilon_B/\varepsilon_F)]^2}$ which
approaches zero faster than $E_+$ as expected (note that in this limit
$Z_-\propto \varepsilon_F/\varepsilon_B$).  In this limit, we find
that the rate $\Gamma_{PM}$ at which the repulsive polaron can form a
molecule is even smaller, like in 3D~\cite{Massignan:2011fk}.
Conservation laws require the creation of an additional particle-hole
pair, and Fermi antisymmetry leads to a suppression by a further
$\varepsilon_F/\varepsilon_B$ by the same argument as in
3D~\cite{PhysRevLett.105.020403}. Thus the rate of the resulting
three-body process is expected to approach zero at least as fast as
$\Gamma_{PM}/\varepsilon_F \sim (\varepsilon_F/\varepsilon_B)^2$.
Eventually, when the attractive polaron is no longer the ground state
(e.g. above $\varepsilon_b/\varepsilon_F\approx10$ for equal masses
\cite{Parish:2011vn}), the pole expansion (\ref{eq:pole}) breaks down
for the attractive polaron and the repulsive polaron will
predominantly decay into the molecular ground state.

For decreasing $\varepsilon_B/\varepsilon_F$, the energy $E_+$
initially steadily increases until it eventually exceeds
$\varepsilon_F$ (see Fig.~\ref{fig:energy}). The interaction at which
this occurs depends sensitively on the mass ratio: For equal masses,
the critical interaction $(\varepsilon_B/\varepsilon_F)_c \simeq 2.7$,
while for $m_\uparrow/m_\downarrow \to \infty$ we have
$(\varepsilon_B/\varepsilon_F)_c \to \infty$, and for
$m_\uparrow/m_\downarrow = 0$ we have $(\varepsilon_B/\varepsilon_F)_c
= 0$. Note that our result in the latter case is consistent with the
exact solution~\cite{Mahan}, $E_+= 
\int_0^{\varepsilon_F} \frac{dy}\pi \cot^{-1}
[\frac1\pi\log\frac{\varepsilon_B}y]$ and in fact we see in
Fig.~\ref{fig:energy} that $E_+$ is quite close to the exact energy,
apart from where the exact energy is non-analytic near
$\varepsilon_B/\varepsilon_F = 0$. Indeed, we expect our variational
approximation \eqref{polaronwavefunction} to be least accurate for the
case of an infinitely massive impurity, where higher order
particle-hole excitations of the Fermi sea are expected to be more
important~\cite{PhysRevLett.101.050404}.

For weak attraction, 
$\varepsilon_B/\varepsilon_F \to 0$, we find that the effective mass
of the repulsive polaron diverges (Fig.~\ref{fig:mass}), signalling
that the lowest-energy repulsive polaron has a finite momentum $P^*$
beyond that point. This was also concluded from the shape of the
spectral function calculated in Ref.~\cite{schmidt2011}. From the
inset of Fig.~\ref{fig:mass}, we see that $P^*$ smoothly evolves from
0 to $k_F$ as $\varepsilon_B/\varepsilon_F \to 0$. This behavior is
tied to the fact that, in this limit, the residue $Z_+$ decreases
towards zero and the wave function evolves into a bare impurity at
momentum $p-k_F$ plus one $\uparrow$-particle excited from zero
momentum to $k_F$.  This limit mirrors the repulsive-polaron solution
for infinite impurity mass, where one removes a $\uparrow$-particle
from the bound state formed from the zero-momentum state and places it
at the Fermi surface.  In terms of the wave function
\eqref{polaronwavefunction}, the final state at
$\varepsilon_B/\varepsilon_F = 0$ corresponds to $\alpha_0^{(\ve{p})}
= 0$ and $\alpha_{\ve{k}\ve{q}}^{(\ve{p})} = \delta_{\ve{q},\ve{0}}
\delta_{\ve{k},-k_F\hat{{\mathbf p}}}$, which is orthogonal to the
attractive polaron (with $Z_-=1$).  Thus, in this limit, the energy is
minimised when $p=k_F$, giving $E_+ = \varepsilon_F$. Contrast this
with the $p=0$ state, where $E_+ =
\varepsilon_F(1+m_\uparrow/m_\downarrow)$.  Both these states are
eigenstates of the non-interacting Hamiltonian and thus the decay rate
of the repulsive polaron approaches zero as
$\varepsilon_B/\varepsilon_F \to 0$ (see Fig.~\ref{fig:decay}). Note
that when $m_+^*<0$, the $p=0$ state may excite a particle out of the
Fermi sea to gain momentum $P^*$. By evaluating the available phase
space close to the divergence of $m_+^*$, the rate at which this
process occurs is found to go as $[E_+(|{\bf p}|=0)-E_+(|{\bf
  p}|=P^*)]^{5/4}$, which implies that the $p=0$ state is metastable
with respect to the finite momentum repulsive polaron in this limit.

Similar finite-momentum excitations are encountered in the
corresponding impurity problem in 1D. The Bethe ansatz solution for
attractive interactions and equal masses exhibits an excited
``broken-pair'' state with negative effective
mass~\cite{McGuire1966}. Here, the excitation's lowest energy is
always at momentum $k_F$ and equals $\varepsilon_F$. By contrast, the
energy at $\ve{p} = 0$ exceeds $\varepsilon_F$, with a maximum of
$2\varepsilon_F$ when the interactions are switched off, like in
2D. The Bethe ansatz solutions are, of course, exact eigenstates of
the 1D system and thus the decay rates for these excitations are
zero. However, in 2D, the repulsive polaron is, at best, metastable
outside of the limits $\varepsilon_B/\varepsilon_F \to 0$ and
$\varepsilon_B/\varepsilon_F \to \infty$, and indeed we see in
Fig.~\ref{fig:decay} that the decay rates are sizeable in the regime
$1\lesssim \varepsilon_B/\varepsilon_F \lesssim 10$ for a range of
mass ratios.

\section{Itinerant ferromagnetism}
To examine how this impacts the existence of itinerant ferromagnetism
in 2D, we start by assuming the presence of spin-polarized domains and
then assessing the stability of these. In order to have stable,
well-defined $\uparrow$ and $\downarrow$ domains, we require there to
be an energy cost for transporting particles across the
$\uparrow$-$\downarrow$ interface.  Thus, we require a repulsive
polaron formed from a $\downarrow$ ($\uparrow$) impurity to have
energy greater than the Fermi energy of the $\downarrow$ ($\uparrow$)
domain. Assuming mechanical equilibrium, where the pressures of each
domain are equal, we then obtain the stability condition $E_+ >
\varepsilon_F^\uparrow \sqrt{m_\uparrow/m_\downarrow}$ ($E_+ > \varepsilon_F^\downarrow
\sqrt{m_\downarrow/m_\uparrow}$).  For equal masses, this simply gives
$E_+ >\varepsilon_F$, which corresponds to
$\varepsilon_B/\varepsilon_F < 2.7$ as described earlier.  At first
sight, this might suggest that ferromagnetism can exist in this
regime; however we see in Fig.~\ref{fig:decay} that the decay rate
$\Gamma$ is large near $\varepsilon_B/\varepsilon_F = 2.7$ and amounts
to a significant fraction of the Fermi energy. The large $\Gamma$
corresponds to a large uncertainty in the energy of the repulsive
polaron, thus allowing particles to tunnel across the interface and
destabilize the domains even when $E_+ >\varepsilon_F$. Once across,
the particles quickly decay into attractive polarons, e.g.\ for the
typical densities used in 2D
experiments~\cite{PhysRevLett.106.105301}, a repulsive polaron with
$\Gamma \sim 0.1\varepsilon_F$ near $\varepsilon_B/\varepsilon_F\sim
1.3$ only has a lifetime of order 1ms.  Of course, the decay rate is
eventually suppressed once $\varepsilon_B/\varepsilon_F \to 0$,
leading one to suspect that ferromagnetism becomes possible in this
limit. However, $Z_+$ is also heavily suppressed as $Z_- \to 1$, which
means that the particles will mostly tunnel directly into the
attractive polaron state (the probability of tunneling into each
polaron state is proportional to $Z_\pm$).  Thus, our results indicate
that saturated ferromagnetism is highly unstable in 2D.

Another route to determining the existence of itinerant ferromagnetism
is to start with a uniform 50/50 mixture of spins and then compare the
formation rate of spin domains with the rate of decay into
$\uparrow$-$\downarrow$ pairs/dimers. In 3D, previous theoretical
studies~\cite{Pekker2011,Zhang2011} have shown that the decay into
pairs always dominates, thus ruling out the spontaneous formation of
ferromagnetic domains. Our stability argument, however, places a
stronger restriction on ferromagnetism in 2D, namely that saturated
ferromagnetism cannot exist even if fully-polarized spin
domains were to be artificially engineered.

A similar scenario holds for unequal masses.  In
Figs.~\ref{fig:energy}-\ref{fig:decay} we show the properties of a
$^{40}$K [$^6$Li] impurity immersed in a $^6$Li [$^{40}$K] Fermi
sea. In this case, we find the domain stability condition to be
satisfied for $\varepsilon_B/\varepsilon_{F,\text{Li}}<5.7$.  However,
again we find that the decay rate is prohibitively large and phase
separation is not favored. Our results are derived under the
assumption that the system is correctly described by the Hamiltonian
(\ref{Hamiltonian}). However, Feshbach resonances in this
heteronuclear system are narrow \cite{PhysRevLett.100.053201} which
changes the two-particle scattering properties. Studying this problem
in detail is beyond the scope of this letter, but the enhanced
closed-channel character close to a narrow resonance decreases the
two-body decay rate of the repulsive polaron~\cite{Innsbruck}. While
in general the energy will also be lower, there may be an intermediate
regime of parameter space which favors phase separation.

Our arguments may be extended to realistic experimental conditions
where the atoms are subjected to a weak trapping potential in the 2D
plane. The stability of the spin polarized domains relies on
mechanical equilibrium at the $\uparrow$-$\downarrow$ interface. This
constraint is local and thus the above considerations are also valid
in the trapped system provided the trap changes smoothly enough that
the local density approximation holds. Additionally, the atoms are
confined to quasi-2D by a strong transverse field characterized by the
frequency $\omega_z$. For the equal-mass case studied in
Ref.~\cite{PhysRevLett.106.105301} with experimental parameters
$\omega_z=2\pi\times80$ kHz and $\varepsilon_F=2\pi\times9$ kHz, we
have $\varepsilon_B\sim\varepsilon_F\ll\omega_z$ in the regime of
interest for the question of possible ferromagnetism. Thus, we
conclude that only the lowest transverse mode is occupied and the 2D
approximation considered in this paper is valid.

\section{Concluding remarks}
We now turn to the question of the accuracy of the variational
wavefunction~(\ref{polaronwavefunction}). In 3D, the energy and
effective mass of the ground-state quasiparticle have been measured in
Refs.~\cite{PhysRevLett.102.230402} and~\cite{PhysRevLett.103.170402},
respectively, and good agreement with the variational
calculation~\cite{PhysRevA.74.063628, Combescot:2007bh} was obtained,
even in the limit of strong interactions. The approach has been
further validated by agreement with Monte Carlo
simulations~\cite{PhysRevB.77.125101}, while
Ref.~\cite{PhysRevLett.101.050404} has argued that the accuracy of the
wavefunction arises from a nearly perfect destructive interference of
the contributions of states dressed by more than one particle-hole
pair.  This latter argument does not depend on dimension and indeed it
can be shown that the ground state energy from the variational
approach agrees well with the exact Bethe Ansatz solution in
1D~\cite{1Dpolaron}.  While the variational wavefunction allows an
accurate calculation of ground state properties of the system, it is,
perhaps, less intuitive that this should be the case for a metastable
polaron. Nevertheless, recent experimental and theoretical
studies~\cite{Innsbruck} of a $^{40}$K-$^{6}$Li mixture in 3D have
demonstrated that not only is the energy and residue of the repulsive
polaron correctly captured in the equivalent diagrammatic approach,
the decay rate is also well approximated by a technique similar to the
one which led to Eq.~(\ref{gamma}).

To conclude, we have calculated the quasiparticle properties of the
repulsive polaron. We find that in the limit of weak
attractive interactions the
lowest lying repulsive polaron has finite momentum. These properties
may be observed by radiofrequency spectroscopy~\cite{schmidt2011}.
However, the fast decay and small residue of these quasiparticles in
the regime of strong repulsion indicates that spin domains will be
unstable and precludes itinerant ferromagnetism in the same manner as
in the 3D experiment \cite{sanner2011}. Our results are also
applicable to a bosonic impurity immersed in a Fermi sea and thus have
implications for phase separation in 2D Bose-Fermi mixtures.

\acknowledgments
  We are grateful to P. Massignan, S. Baur, N. Cooper, and M. K\"ohl
  for fruitful discussions.  JL acknowledges support from a Marie
  Curie Intra European grant within the 7th European Community
  Framework Programme. MMP acknowledges support from the EPSRC under
  Grant No.\ EP/H00369X/1.
  
\bibliography{Ref2D}
\bibliographystyle{eplbib}
  
\end{document}